\documentclass[a4paper,12pt]{article}
\pdfoutput=1
\usepackage{graphicx,rotating,hyperref,slashed,amsmath,xcolor,amssymb,amsfonts,expdlist,colortbl,cite,youngtab}
\usepackage{multirow}
\usepackage{booktabs}    
\usepackage{calc}        
\makeatletter
\hypersetup{colorlinks,bookmarksopen,bookmarksnumbered,
linkcolor=blus,pdfstartview=FitH,urlcolor=rossos,citecolor=verde}

\def\lsim{\mathrel{\rlap{\lower3pt\hbox{\hskip0pt$\sim$}}
   \raise1pt\hbox{$<$}}}         
\def\gsim{\mathrel{\rlap{\lower4pt\hbox{\hskip1pt$\sim$}}
   \raise1pt\hbox{$>$}}}         

 \newcommand{\sfootnote}[1]{}
\definecolor{bluc}{cmyk}{1,1,0,0.1}
\definecolor{rossoCP3}{cmyk}{0,.88,.77,.40}
\definecolor{rosso}{cmyk}{0,1,1,0.4}
\definecolor{rossos}{cmyk}{0,1,1,0.55}
\definecolor{rossoc}{cmyk}{0,1,1,0.2}
\definecolor{verdes}{cmyk}{0.92,0,0.59,0.4}

\hypersetup{colorlinks, bookmarksopen, bookmarksnumbered,
citecolor=verdes, linkcolor=bluc, pdfstartview=FitH, urlcolor=rossos}

\newcommand{\mio}[1]{}

\newcommand{\fig}[1]{~\ref{fig:#1}}

\definecolor{Gray}{gray}{0.95}

\newcommand{\sfrac}[2]{#1/#2}

\def\bac{\begin{array} {c}}
\def\bacc{\begin{array} {cc}}
\def\baccc{\begin{array} {ccc}}
\def\bacccc{\begin{array} {cccc}}
\def\ea{\end{array}}

\usepackage{multicol}
\usepackage{color}
\definecolor{rosso}{cmyk}{0,1,1,0.4}
\definecolor{rossos}{cmyk}{0,1,1,0.55}
\definecolor{rossoc}{cmyk}{0,1,1,0.2}
\definecolor{blu}{cmyk}{1,1,0,0.3}
\definecolor{blus}{cmyk}{1,1,0,0.6}
\definecolor{bluc}{cmyk}{1,1,0,0.1}
\definecolor{verde}{cmyk}{0.92,0,0.59,0.25}
\definecolor{verdec}{cmyk}{0.92,0,0.59,0.15}
\definecolor{verdes}{cmyk}{0.92,0,0.59,0.4}

\newcommand{\riga}[1]{\noalign{\hbox{\parbox{\textwidth}{#1}}}\nonumber}

\newcommand{\GeV}{\,{\rm GeV}}
\newcommand{\TeV}{\,{\rm TeV}}

\def\circa#1{\,\raise.3ex\hbox{$#1$\kern-.75em\lower1ex\hbox{$\sim$}}\,}

\newcommand{\beq}{\begin{equation}}
\newcommand{\eeq}{\end{equation}}

\newcommand{\bea}{\begin{eqnarray}}
\newcommand{\eea}{\end{eqnarray}}
\newcommand{\be}{\begin{equation}}
\newcommand{\ee}{\end{equation}}
\font\tenrsfs=rsfs10 at 12pt
\font\sevenrsfs=rsfs7 at 10 pt
\font\fiversfs=rsfs5
\newfam\rsfsfam
\textfont\rsfsfam=\tenrsfs
\scriptfont\rsfsfam=\sevenrsfs
\scriptscriptfont\rsfsfam=\fiversfs
\def\mathscr#1{{\fam\rsfsfam\relax#1}}

\def\circa#1{\,\raise.3ex\hbox{$#1$\kern-.75em\lower1ex\hbox{$\sim$}}\,}
\makeatletter

\def\hhref#1{\href{http://arxiv.org/abs/#1}{arXiv:#1}} 

\newcommand{\doi}[1]{\href{http://dx.doi.org/#1}{[doi]}}

\renewenvironment{thebibliography}[1]
     {\begin{multicols}{2}[\section*{\refname}]%
      \@mkboth{\MakeUppercase\refname}{\MakeUppercase\refname}%
      \list{\@biblabel{\@arabic\c@enumiv}}%
           {\settowidth\labelwidth{\@biblabel{#1}}%
            \leftmargin\labelwidth
            \advance\leftmargin\labelsep
            \@openbib@code
            \usecounter{enumiv}%
            \let\p@enumiv\@empty
            \renewcommand\theenumiv{\@arabic\c@enumiv}}%
      \sloppy
      \clubpenalty4000
      \@clubpenalty \clubpenalty
      \widowpenalty4000%
      \sfcode`\.\@m}
     {\def\@noitemerr
       {\@latex@warning{Empty `thebibliography' environment}}%
      \endlist\end{multicols}}

\font\ital=cmu10 
\def\hhref#1{\href{http://arxiv.org/abs/#1}{arXiv:#1}}
\usepackage{xstring} 
\newcommand{\hhrefq}[1]{\IfSubStr{#1}{:}{\href{http://inspirehep.net/search?ln=en&ln=en&p=#1&of=hb&action_search=Search&sf=&so=d&rm=&rg=25&sc=0}{InSpires:#1}}{\hhref{#1}}}

\def\art{\@ifnextchar[{\eart}{\oart}}
\def\eart[#1]#2#3#4#5#6{{\rm #2}, {\em #3 \bf #4} {\rm (#6) #5} ({\em #1})}
\def\article{\@ifnextchar[{\earticle}{\oarticle}}
\def\oarticle#1#2#3#4#5#6{{\rm #1}, {\ital ``#6''}, {\rm #2 #3 (#5) #4}}
\def\earticle[#1]#2#3#4#5#6#7{{\rm #2}, {\ital ``#7''}, {\rm #3 #4 (#6) #5}  [\hhrefq{#1}]}
\def\hepart[#1]#2{{\rm #2, \sl#1}}
\def\heparticle[#1]#2#3{#2, {\ital ``#3''} [\hhrefq{#1}]}

\newcounter{alphaequation}[equation]
\def\thealphaequation{\theequation\hbox to
0.6em{\hfil\alph{alphaequation}\hfil}}
\def\eqnsystem#1{
\def\@eqnnum{{\rm (\thealphaequation)}}
\def\@@eqncr{\let\@tempa\relax \ifcase\@eqcnt \def\@tempa{& & &} \or
  \def\@tempa{& &}\or \def\@tempa{&}\fi\@tempa
  \if@eqnsw\@eqnnum\refstepcounter{alphaequation}\fi
\global\@eqnswtrue\global\@eqcnt=0\cr}
\refstepcounter{equation} \let\@currentlabel\theequation \def\@tempb{#1}
\ifx\@tempb\empty\else\label{#1}\fi
\refstepcounter{alphaequation}
\let\@currentlabel\thealphaequation
\global\@eqnswtrue\global\@eqcnt=0 \tabskip\@centering\let\\=\@eqncr
$$\halign to \displaywidth\bgroup \@eqnsel\hskip\@centering
$\displaystyle\tabskip\z@{##}$&\global\@eqcnt\@ne
\hskip2\arraycolsep\hfil${##}$\hfil& \global\@eqcnt\tw@\hskip2\arraycolsep
$\displaystyle\tabskip\z@{##}$\hfil
\tabskip\@centering&\llap{##}\tabskip\z@\cr}
\def\endeqnsystem{\@@eqncr\egroup$$\global\@ignoretrue} \makeatother

\newcommand{\SU}{\,{\rm SU}}

\newcommand{\U}{\,{\rm U}}

\definecolor{fiorentina}{rgb}{.5,0,.5}

\allowdisplaybreaks

\begin{document}
\centerline{CERN-{TH}-2017-173 \hfill  CP3-Origins-2017-029  \hfill IFUP-TH/2017 \hfill IPPP/17/61}

\vspace{1truecm}

\begin{center}
\boldmath

\mbox{{\textbf{\LARGE\color{rossoCP3} Asymptotically Safe Standard Model Extensions?}}}
\unboldmath

\bigskip\bigskip\bigskip

\large
{\bf Giulio Maria Pelaggi}$^a$,
{\bf Alexis D. Plascencia}$^b$,
{\bf Alberto Salvio}$^{c}$, 
{\bf Francesco Sannino}$^{d,c}$, 
{\bf Juri Smirnov}$^{e}$, 
{\bf Alessandro Strumia}$^{a,c}$
 \\[8mm]
{\it $^a$ Dipartimento di Fisica dell'Universit{\`a} di Pisa and INFN, Italy}\\[1mm]
{\it $^b$ Institute for Particle Physics Phenomenology, Department of Physics, Durham University, Durham DH1 3LE, United Kingdom}\\[1mm]
{\it $^c$  Theoretical Physics Department, CERN, 1211 Geneva 23, Switzerland}\\[1mm]
{\it $^d$ CP$^3$-Origins and Danish IAS, University of Southern Denmark, 
Denmark}\\[1mm]
{\it $^e$ INFN and Department of Physics and Astronomy\\ University of Florence,
Via G. Sansone 1, 50019 Sesto Fiorentino, Italy}\\[1mm]

\bigskip\bigskip\bigskip

\thispagestyle{empty}\large
{\bf\color{blus} Abstract}
\begin{quote}
We consider theories with a large number $N_F$ of charged fermions and
compute the renormalisation group equations for the gauge, Yukawa and quartic couplings resummed at leading order in $1/N_F$.
We construct extensions of the Standard Model where SU(2) and/or SU(3) are asymptotically safe.
When the same procedure is applied to the Abelian U(1) factor, we find that the 
Higgs quartic can not be made asymptotically safe and stay perturbative at the same time.
\end{quote}
\thispagestyle{empty}
\end{center}

\setcounter{page}{1}
\setcounter{footnote}{0}

\newpage

\tableofcontents


\section{Introduction}

The  Large Hadron Collider (LHC) data collected at  $\sqrt{s}=13\TeV$ are in line with the Standard Model (SM) predictions
and provide strong bounds on SM extensions, in particular on those that were introduced to tame the quadratic sensitivity of the Higgs mass operator on the scale of new physics, known as natural extensions.   The time is therefore ripe to explore alternative approaches to naturalness and even better, new guiding principles, that can help selecting a more fundamental theory of Nature. 
  
Weak-scale extensions of the SM valid up to infinite energy bypass the issue of quadratically divergent corrections to the Higgs mass~\cite{1412.2769} (we will ignore gravity, as an extension of Einstein gravity can have this property~\cite{1403.4226,1705.03896}).
Extensions of the SM realising total asymptotic freedom can be built by embedding the Abelian U(1)$_Y$ into non-Abelian
gauge groups that explain the observed values of the hypercharges.
Natural possibilities, where the extended gauge group is broken around the electro-weak scale~\cite{1412.2769},
have been proposed based on the groups
$\SU(2)_L\otimes\SU(2)_R\otimes\SU(4)_c$ and especially
$\SU(3)_L\otimes\SU(3)_R\otimes\SU(3)_c$~\cite{1507.06848}.
However, a fine-tuning at the $\%$ level is needed  in order to make the
extra vectors above present bounds, such as $M_{W_R}>2.5\TeV$~\cite{1707.01303}. 

This is one of the motivations behind the search of alternative fundamental SM extensions that are  asymptotic safe, rather than asymptotically free. In addition, asymptotically safe theories are an intriguing and yet much unexplored possibility. Only recently  the first controllable perturbative example of a gauge-Yukawa theory able to display asymptotic safety in all couplings~\cite{SL} was discovered,
where the Veneziano-Witten limit $N,N_F\gg 1$
has been employed to unquestionably establish the existence of such a scenario. 
Quantum stability of the theory and the determination of the vacuum and potential of the theory were established in \cite{Litim:2015iea}. The original model did not feature gauged scalars nor radiative symmetry breaking and is a vector-like theory.  Gauged scalars and related asymptotically safe conditions were introduced for the first time in \cite{1701.01453} while  chiral gauge theories were investigated in \cite{Molgaard:2016bqf,1701.01453}, and radiative breaking in \cite{Abel:2017ujy}.  Extensions to semi-simple groups with SM-like chiral matter appeared first  in \cite{Esbensen:2015cjw} while semi-simple gauge theories with vector-like fermions appeared in \cite{Bond:2017lnq}.  The Veneziano limit leads to interesting phenomenological applications once spontaneous breaking occurs \cite{1707.06638}.  Another virtue of a controllable perturbative limit is that in these theories the Higgs mass can be naturally lighter than the transition scale $\Lambda$, given that non-perturbative corrections are  exponentially suppressed, $\delta M_h^2 \sim \Lambda^2 e^{-{\cal O}(1)/\alpha}$~\cite{1701.01453}. 

 Supersymmetric asymptotically safe quantum field theories, 
where exact non-perturbative results have been established  in \cite{Intriligator:2015xxa,Bajc:2016efj}, are also an intriguing possibility albeit presumably not of natural type because of the tension with the Large Electron-Positron Collider (LEP) and LHC bounds~\cite{SUSY}.

In this work we depart from the Veneziano limit and supersymmetric extensions by taking another interesting  theoretical limit  that can help taming the ultimate fate of one or all of the SM U(1)$_Y$, $\SU(2)_L$ and $\SU(3)_c$  gauge factors.
This makes use of the large number of flavour expansion discussed in~\cite{1006.2119,1011.5917,1311.5268}. 
In the presence of $N_F\gg 1$ extra fermions one can resum corrections at leading order in $1/N_F$~\cite{PalanquesMestre:1983zy,Gracey:1996he}.  As we will review in section~\ref{Gauge functions}, these fixed points occur at predicted nonperturbative values of the product of their gauge couplings times the associated large number of extra flavours.  
Unsuccessful attempts of constructing perturbative asymptotically safe extensions of the SM appeared in \cite{1702.01727}. Here both the hypercharge and scalar quartics were still under the spell of Landau poles, and in any event the constructions depart from the rigorous limit of \cite{SL}.
 
The renormalization group equations (RGEs) for the Yukawa couplings get modified by the resummation~\cite{1712.06859}. 
We here compute how the RGE for the quartic Higgs coupling gets modified
and apply our results to the case of the Standard Model.

 The paper is structured as follows.
In section~\ref{beta} we review how the introduction of
many extra fermions allows for an ultraviolet interacting fixed point for the gauge couplings,
and compute how the RGE for Yukawa and scalar quartic couplings get modified.
In section~\ref{minimal} we show that this allows to make the SU(2)$_L$ and/or SU(3)$_c$ factors of the SM gauge group 
asymptotically safe.  However we find that U(1) cannot be made asymptotically safe in a controlled regime.
We offer our conclusions  in section~\ref{concl}.

\begin{figure}[t]
$$\includegraphics[width=0.9\textwidth]{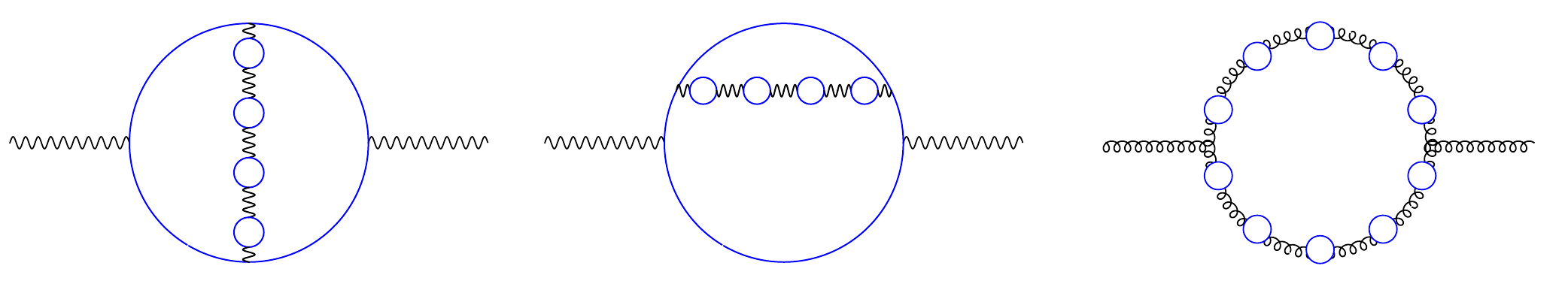} $$
\caption{\em   Feynman diagram topologies that dominate at leading order in $1/N_F$ (we do not show tadpoles).\label{fig:SerieBolle}}
\end{figure}%

\section{Fixed points in the large number of flavours}\label{Gauge functions}\label{beta}     
We now discuss the fate of gauge theories at short distance in the presence of a large number of vector-like fermions ($N_F\gg1$). We do this by first summarising the associated $\beta$-functions resummed at leading order in  $1/ N_F$. We denote with  $\alpha_i \equiv g_i^2/4\pi$ for $i=Y, 2, 3$ the SM gauge couplings (we define $g_Y$  as the hypercharge gauge coupling in the normalization where the Higgs $H$ has $|Y|=1/2$).

\subsection{Resummed gauge $\beta$-functions}
We conveniently write the gauge $\beta$-functions as
\be   \frac{\partial\alpha_i}{\partial \ln \mu} = \beta_{\alpha_i} =  \beta_{\alpha_i}^{\rm SM} +\beta_{\alpha_i}^{\rm extra}\  ,  \ee  
where $\beta_{\alpha_i}^{\rm SM}$ are the perturbative SM contribution:
at one-loop 
$\beta_{\alpha_i}^{\rm SM} = b^{\rm SM}_i \alpha_i^2/2\pi$, with $b^{\rm SM}_Y = 41/6$, $b^{\rm SM}_2 =  - 19/6$ and $b_3^{\rm SM} = -  7$.
The contribution of the $N_F \gg 1 $ extra fermions can be written as their one-loop contribution plus their resummation
at leading order in $1/N_F$  (see \cite{1006.2119,1707.02942}  and reference therein): 
\be \label{eq:betaextra}
\beta_{\alpha_i}^{\rm extra} = \frac{\alpha_i^2 }{2\pi} {\Delta b_i}  +  \frac{\alpha_i^2 }{3\pi}  F_i( \Delta b_i\frac{\alpha_i}{4\pi})  . \ee
The one-loop coefficients are well known:
for Dirac fermions in the representation $R_i$ with dimension $D_{R_i}$  and Dynkin index $S_{R_i}$ they are given by
\be \Delta b_Y  =  \frac43 Y^2 N_F D_{R_2} D_{R_3}, \qquad  \Delta b_2 = \frac{4}{3} N_F S_{R_2}D_{R_3}, \qquad  \Delta b_3 = \frac{4}{3} N_F S_{R_3} D_{R_2}  .  \label{DeltabExpr}\ee 
\begin{figure}[t]
$$\includegraphics[width=0.6\textwidth]{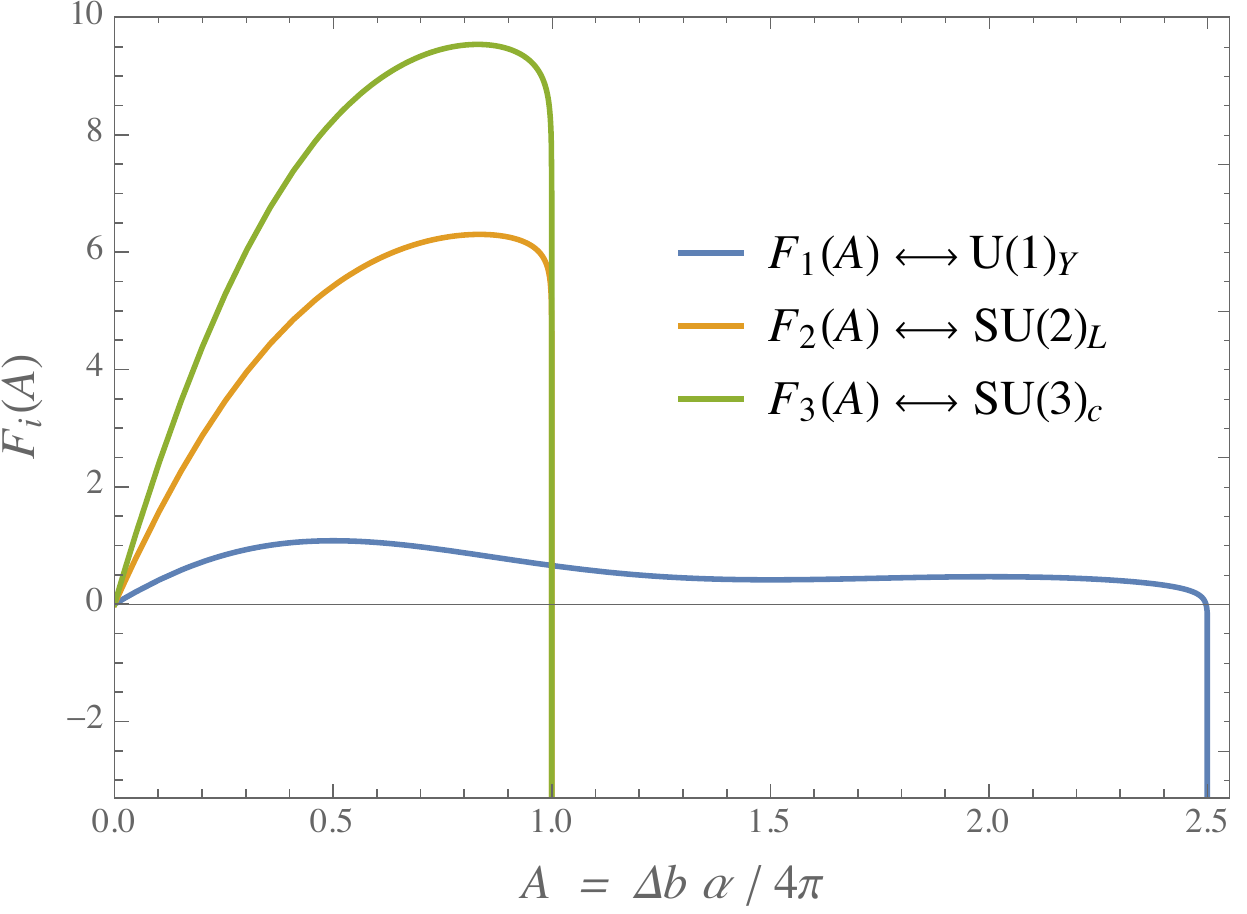} $$
 \caption{\em The functions giving the resummed gauge $\beta$-functions in eq.s~(\ref{FYH1i}), (\ref{I1}) and (\ref{I2}).
\label{fig:FYH1i}
}
\end{figure}%
We write the resumed contributions in the simpler limit where only one of the $\Delta b_i$ is non-vanishing,
in order to the neglect the mixed contributions. 
At leading order in $N_F\gg 1$, the result is dominated by the Feynman diagrams in fig.\fig{SerieBolle}
(extra diagrams are present for non-Abelian groups).
Their  resummation gives the  functions $F_i(A)$  
 \bea F_1(A)  &\equiv& 2 \int_0^{A} I_1(x) dx, \qquad   F_n(A) \equiv  \int_0^{A} I_1(x) I_n(x)  \quad\hbox{ for }n=2,3 \label{FYH1i}  \\ 
\riga{where}\\ 
  I_1(x) &\equiv& \frac{(1+x)(2x-1)^2 (2x-3)^2 \sin^3(\pi x) \Gamma(x-1)^2 \Gamma(-2x)}{\pi^3 (x-2)}, \label{I1} \\ 
  I_n(x) &\equiv& \frac{n^2 -1}{2 n} +\frac{(20- 43 x +32 x^2 -14 x^3 +4x^4) n}{2(2x-1)(2x-3)(1-x^2)} \label{I2}.\eea 
 The $F_i$ functions are plotted in fig.~\ref{fig:FYH1i}. Note that $F_1$ ($F_{2,3}$) has a logarithmic singularity at $A=5/2$ ($A=1$)
 \be F_1(A) \stackrel{A \to 5/2}\simeq \frac{14}{15 \pi ^2}\ln \left(1-\frac{2 A}{5}\right)+ 0.611+\cdots  ,\qquad
 F_n(A) \stackrel{A\to 1}{\simeq}  \frac{n}{8}  \ln \left(1-A\right) + \cdots \ee
 where $\cdots$ in the latter expression
 represents  contributions which remain finite at the singularity.
 As clear from fig.~\ref{fig:FYH1i}, the locations of these singularities  coincide to a very good approximation with the points 
 where the $\beta$-functions vanish, leading to the fixed points 
\beq \label{eq:alpha*}
\alpha_{2,3}^* = \frac{4\pi}{\Delta b_{2,3}},\qquad \alpha_Y^* = \frac{10\pi}{\Delta b_Y}.\eeq
Such fixed-point values reproduce the observed values of the SM gauge couplings renormalized at a few TeV
for $\Delta b_Y \approx 1800$, $\Delta b_2 \approx 400$, $\Delta b_3 \approx 150$:
 higher values of $\Delta b_i$ are thereby not allowed.

\smallskip
   
A word of caution is in order here:
the fixed point resulting from the large-$N_F$ resummation is not
on the same rigorous  footing as the perturbative fixed points arising in the limit of large $N,N_F$~\cite{SL},
or the supersymmetric fixed points~\cite{Intriligator:2015xxa,Bajc:2016efj}. 
The physical meaning of the logarithmic singularity needs to be investigated further, especially because extra singular behaviours emerge  at sub-leading orders in $N_F$~\cite{1006.2119}. Of course, the important issue is whether the first singularity is unaffected by the emergence of new singularities that might very well imply the existence of different physical branches not linked to the original one as argued in~\cite{1006.2119}. In fact, we can start to understand how the first UV fixed point starts to emerge within perturbation theory \cite{1011.5917}. 
Singularities in beta functions are not a pathology of the theory as  the well known exact supersymmetric beta functions show \cite{Novikov:1983uc}. In fact using alternative large number of colours limits one can even map these supersymmetric beta functions in the one of one flavour QCD \cite{hep-th/0302163,hep-th/0309252}. Furthermore beta functions are scheme dependent while fixed points are physical.  This means that one can, in principle, find another scheme in which the beta function has a different behaviour while the theory retains the UV fixed point. This again happens in supersymmetric field theories when going from the all-order exact beta function to the exact Wilsonian (holomorphic scheme) one-loop exact beta function, see \cite{hep-th/0207163} for a discussion of the scheme transformations and their impact on their derivation of the beta functions via string theory. Clearly the transformation among the schemes is also singular. 
 
 \smallskip
 
Another possible issue is that
the resummation of a perturbative series can produce meaningless results when the series
is asymptotic, not necessarily convergent (this happens in the SM for the series in the quartic SM Higgs coupling).
In our case the fermionic path integral gives a functional determinant which is an analytic function
of the gauge coupling, such that the final path integral over vectors can be expanded in a convergent series.
The predictions from resummation can be confirmed via lattice simulations for which neither technical nor theoretical impediment exist.  
Furthermore, lattice simulations can test whether fixed points exist also away from the $N_F\gg 1$ limit.
No fixed point was found in QED with one electron, and this computation can be repeated with larger $N_F$.

\begin{figure}[t]
\begin{center}
\includegraphics[width=0.95\textwidth]{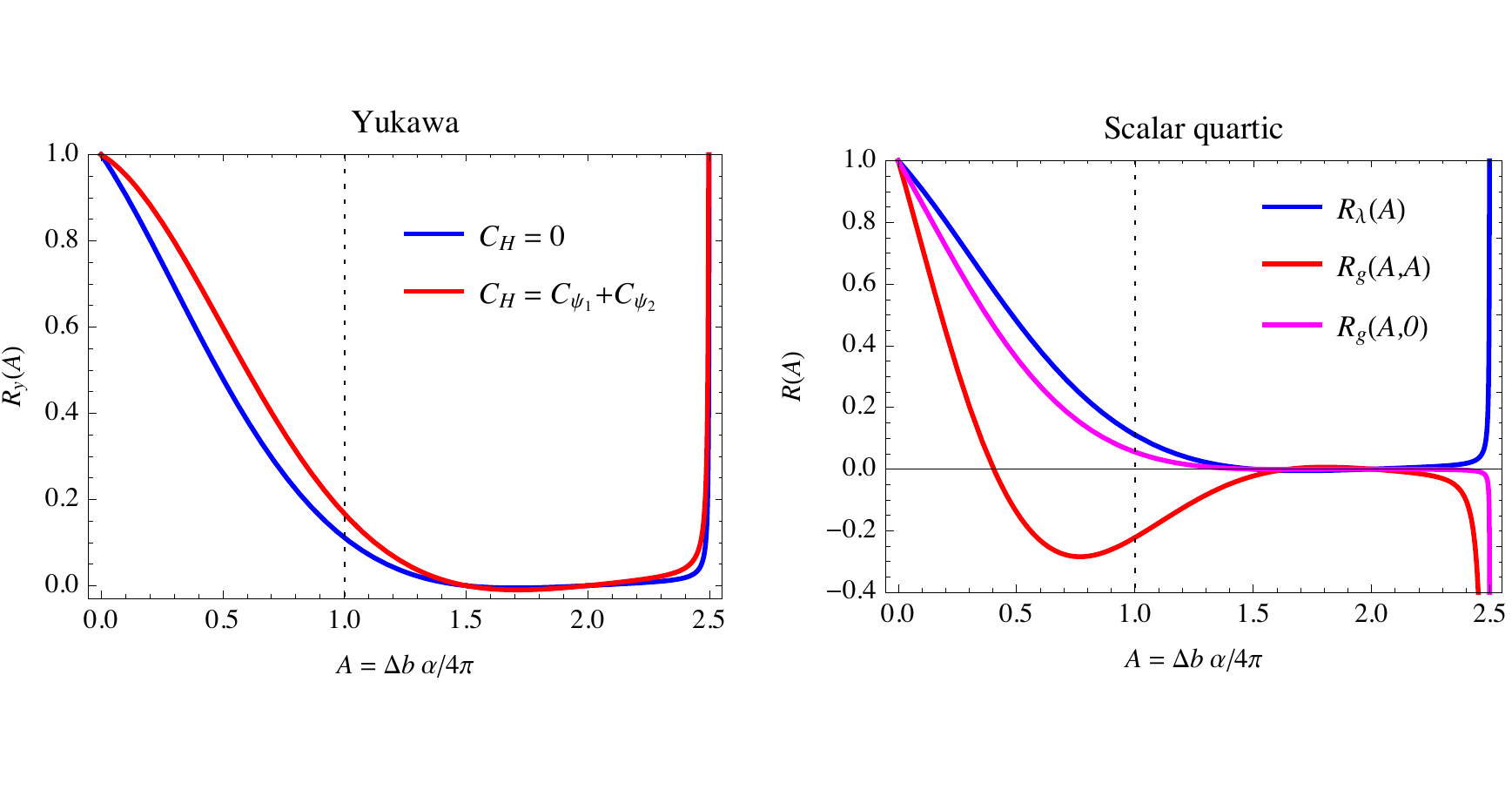} 
\vspace{-1cm}
\caption{\label{Aphi}\em The functions $R_y$ (left) and
$R_\lambda, R_g$ (right) encoding the correction due to the extra fermions in the large-$N_F$ limit to the $\beta$-function of the Yukawa and quartic couplings.  
}
\end{center}
\end{figure}

\subsection{Resummed Yukawa $\beta$-function} \label{YukSec}
As pointed out in~\cite{1712.06859}, one needs to resum corrections to the RGE of Yukawa couplings.
We consider a Yukawa coupling $y\, \psi_1 \psi_2 H + \hbox{h.c.}$ where $\psi_{1,2}$ are Weyl fermions
and $H$ is a scalar field. All particles are in generic representations of the gauge group $G = \prod_i G_i$.
The one-loop $\beta$-function improved by  resumming the gauge propagators can be written as
 \beq \frac{\partial y}{\partial \ln\mu}= \beta_y 
= -3 \frac{y}{4\pi}   \sum_i (C_{\psi_1 i}+ C_{\psi_2 i} )\alpha_i  \times R_y(A_i) + {\cal O}(y^3) \eeq
where $C_{\psi i}$ is the quadratic Casimir of $\psi$ under $G_i$.
The well known ${\cal O}(y^3)$ terms are not shown because not affected by the resummation, encoded
in the function $R_y(A_i)$, equal to $R_y(0) =1 $ in the limit of vanishing $A_i = \Delta b_i\, \alpha_i/4\pi$.
This is computed in appendix~\ref{Ry}, with the result 
\beq R_y(A)= \frac{(3-2 A)^2(2-A) \sin (\pi  A) \Gamma (2-2 A) }{9 \pi A   \Gamma (3-A)^2}
   \left(2 + A\frac{C_H}{C_{\psi_1 }+ C_{\psi_2 } }\right) 
   \eeq
which agrees with the computation in~\cite{1712.06859}.
For a U(1) one has $C_{p} =q_{p}^2$, where $q_p$ is the charge of particle $p$, that satisfies $q_{\psi_1}+q_{\psi_2}+ q_H=0$.
If multiple U(1) are present, the expressions above hold in a basis where they do not mix.


Particularly relevant are the values of $R_y$ close to the fixed point of gauge couplings: $A=1$ for a non-Abelian coupling,
and $A=5/2$ for an Abelian coupling:
\beq R_y(1) = \frac{1}{18}  \left(2 + \frac{C_H}{C_{\psi_1 }+ C_{\psi_2 } }\right) ,\qquad
R_y(A)  \stackrel{A\to 5/2}{\simeq}
   \frac{9 + 10 q_{\psi_1} q_{\psi_2}/(q_{\psi_1}^2+q_{\psi_2}^2)}{270\pi^2 (5/2-A)}   .
   \eeq
The pole at $A \to 5/2$ implies that Yukawa couplings of  fermions charged under an asymptotically safe U(1)
are driven to negligibly small values at large energies.


\subsection{Resummed quartic $\beta$-function}

At leading order in $1/N_F$,  $\beta_{\lambda}$ is given by
\beq 
\frac{\partial \lambda}{\partial \ln \mu} = {\beta}_{\lambda} =  - \frac{ \lambda}{4\pi} \sum_i C_i \alpha_i
R_{\lambda} (A_i) + \sum_{ij} C_{ij} \alpha_i \alpha_j R_{g}(A_i,A_j) + {\cal O}(\lambda^2, \lambda y^2, y^4),
\eeq
 where $C_i$ and $C_{ij}$ are the well known one-loop $\beta$-function coefficients,
and the terms of order $\lambda^2, \lambda y^2, y^4$ are not affected by the resummation
(unless there is a large number $N_F$ of Yukawa couplings; in this case one needs to resum them too).
The values of $C_i$ and $C_{ij}$ in a generic QFT can be found  in~\cite{Machacek:1984zw}.
We compute the functions $R_\lambda(A)$ and $R_g(A)\equiv R_g(A,A)$ in appendix~\ref{Rlambda}, finding
\bea R_g(A,A) &=& \frac{\left[(2 A-3) A \left(H_A-3 H_{1-A}+2 H_{3-2 A}\right)-4 A+3\right] \Gamma (4-2 A)}{18 \Gamma (2-A)^3 \Gamma (A+1)}, \label{Rgeq}\\
  R_g(A,0) & =& \frac{(3-2 A) \Gamma (4-2 A)}{18 \Gamma (2-A)^3 \Gamma (A+1)},\\
  R_\lambda \left(A\right) &=&
   \frac{2 (3-2 A) \Gamma(4-2A)}{9 A(4-2A)  \Gamma (2-A)^3 \Gamma(A)}, \label{Rlambdaeq}
 \eea
with $H_n$ the $n^{\rm th}$ harmonic number.
These functions are plotted in  fig.~\ref{Aphi} and are regular at $A=1$, close
to the fixed point of non-Abelian gauge couplings: $R_\lambda(1) = \sfrac{1}{9}, \, R_g(1,1) = -\sfrac{2}{9}$,
$R_g(1,0)=1/18$.  
On the other hand, they diverge for $A\to 5/2$, close to the fixed point of
Abelian gauge couplings. 
The leading behaviour as $A\to 5/2$ is 
\be R_\lambda(A) \simeq -\frac{2}{135 \pi ^2 \left(A-\sfrac{5}{2}\right)}  + \cdots\, \eeq
and
\beq
 R_g(A,0) \simeq \frac{1}{270 \pi ^2 \left(A-\sfrac{5}{2}\right)}  + \cdots\, , \qquad 
R_g(A,A) \simeq -\frac{1}{108 \pi ^2 \left(A-\sfrac{5}{2}\right)^2} + \cdots \, ,\ee
where the dots represent terms that are regular as $A\to 5/2$. 

%
\section{Asymptotically safe SM extensions}\label{minimal}

\begin{figure}[t]
$$\includegraphics[width=0.47\textwidth]{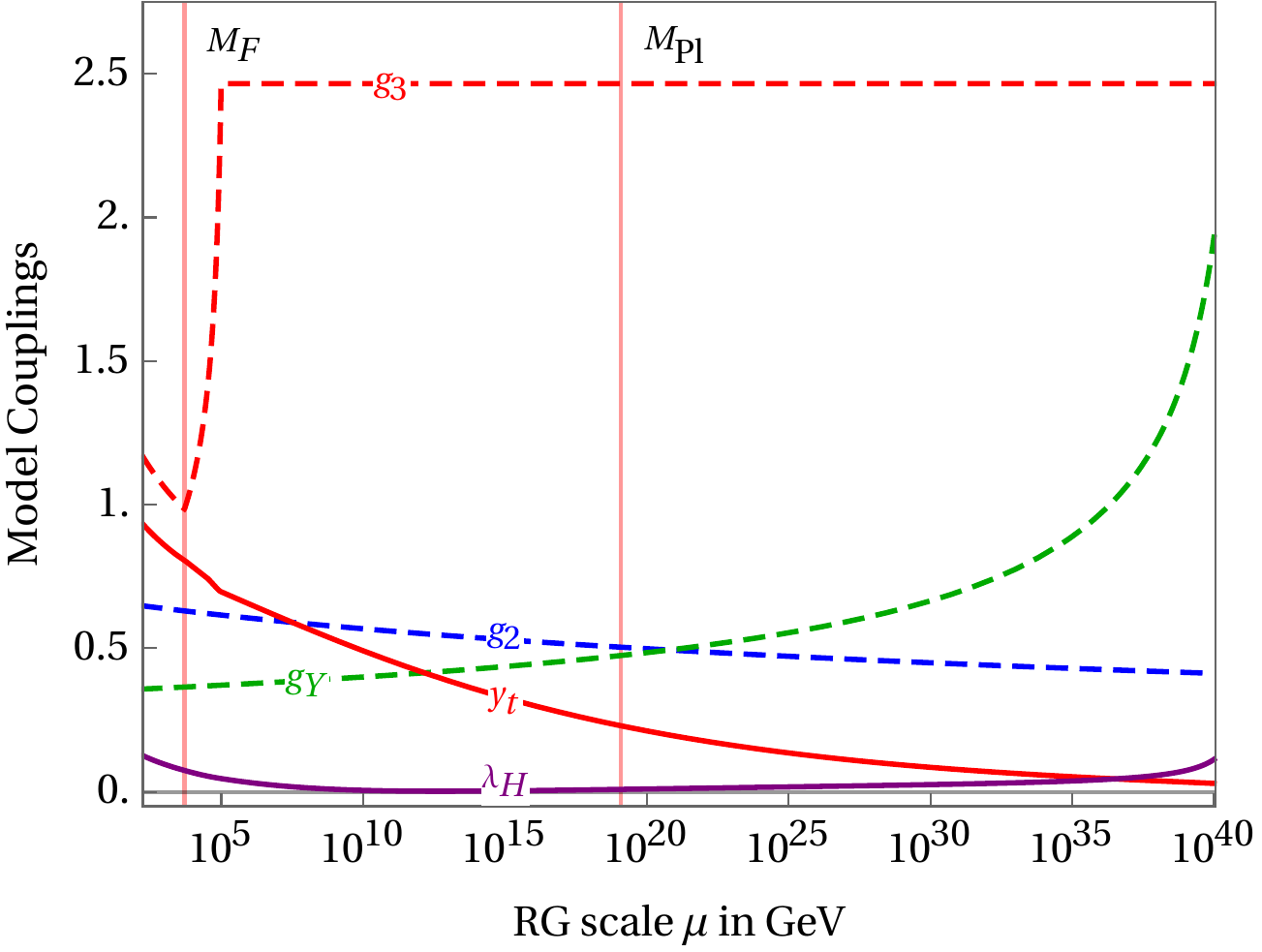} \includegraphics[width=0.47\textwidth]{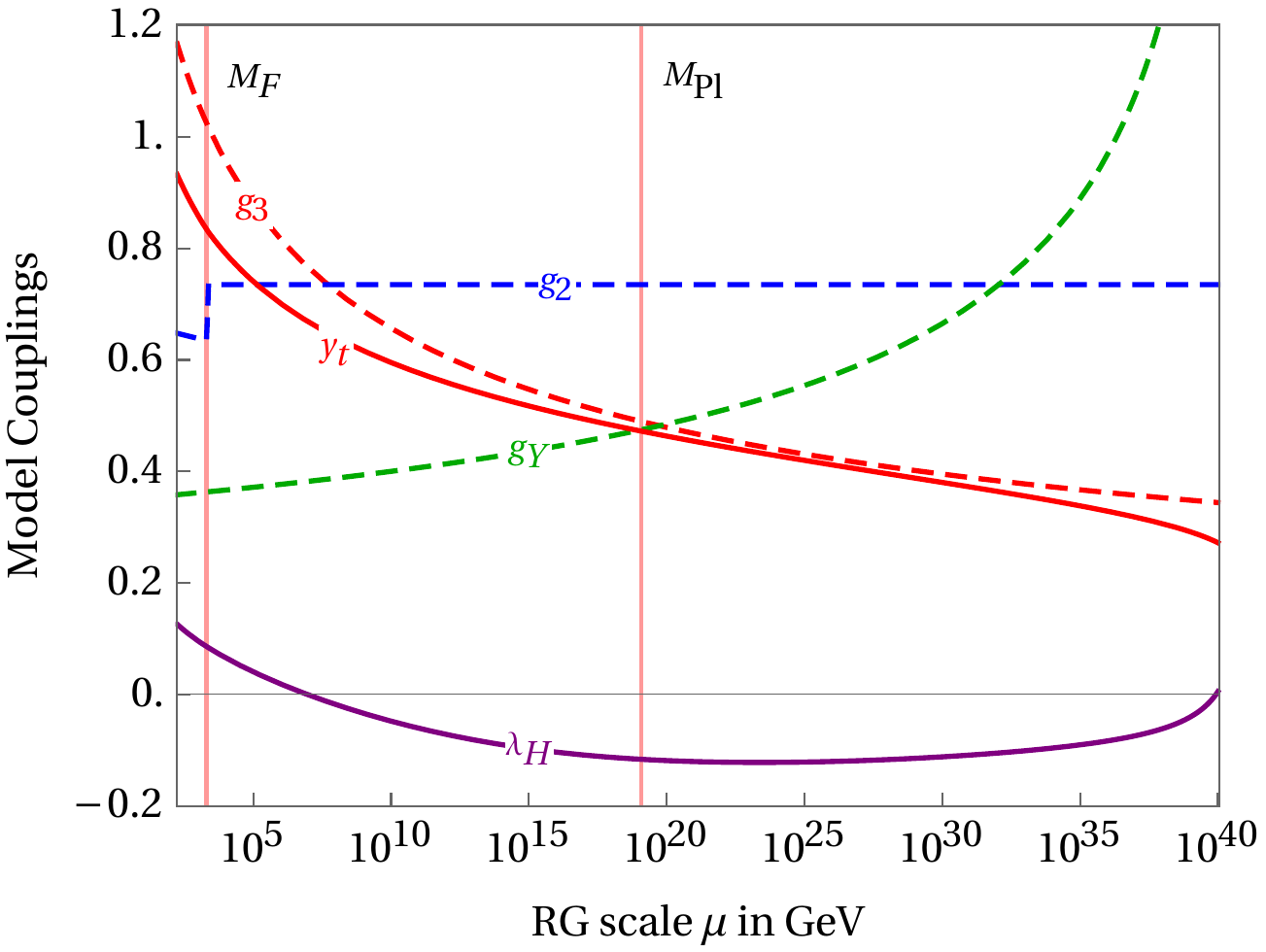}$$
\caption{\em Left panel: running couplings in the SM plus $N_F=13$ $(\Delta b_3=26)$ fermion color octets with mass $M_F\approx 5 \TeV$: $\SU(3)_c$ becomes asymptotically safe; the electroweak vacuum becomes stable. Right panel: running couplings in the SM plus $N_F=220$  $(\Delta b_2=880/3)$  $\SU(2)_L$ fermionic triplets with mass $M_F\approx 2 \TeV$: $\SU(2)_L$ becomes asymptotically safe; the electroweak vacuum becomes unstable. 
\label{fig:GaugeRunning}}
\end{figure}

\subsection{Asymptotically safe $\SU(3)_c$}
We add to the SM $N_F$  Majorana fermion color octets with no weak interactions,
such that $\Delta b_3 = 2N_F$, while $\Delta b_2 = \Delta b_Y=0$.
Then, the resummed one-loop beta functions for the top Yukawa coupling is
\bea
(4\pi)^2 \frac{dy_t}{d\ln\mu} &=& \frac92 y_t^3 - y_t \bigg(8 g_3^2 R_y(A_3)+ \frac94 g_2^2  + \frac{17}{12} g_Y^2\bigg)  
\eea
while the RGE for the Higgs quartic remains as in the SM.
Close to the fixed point $R_y(1)=1/9$, such that the running of $y_t$ is mildly modified.
In our numerical example\footnote{For fermions in the adjoint representation the lower boundary of the safe conformal region has been estimated to be around seven flavours \cite{1709.02354}.  It is consistently and considerably lower than for fermions in the fundamental representation.} in fig.~\fig{GaugeRunning} (left) this has a minor indirect effect on the running of the Higgs quartic.  Nevertheless, by choosing $N_F=13$ (or smaller values) it is possible to make the electroweak vacuum stable. To solve the RG equations we take the central value for the top quark mass from recent measurements performed by ATLAS and CMS, $M_t=172.5 $ GeV \cite{Mtexp}.


\subsection{Asymptotically safe $\SU(2)_L$}
We add to the SM $N_F$  Majorana fermion triplets with no hypercharge and no color,
such that $\Delta b_2 = 4N_F /3$,  while $\Delta b_3 = \Delta b_Y=0$.
Then, the resummed one-loop beta functions for the top Yukawa coupling and for
the Higgs quartic $\lambda_H$
(defined writing  the  tree-level SM potential as $V = -\frac12 M_h^2|H|^2 +  \lambda _H |H|^4$)
are
\bea
(4\pi)^2 \frac{dy_t}{d\ln\mu} &=& \frac92 y_t^3 - y_t \bigg(8 g_3^2 + \frac94 g_2^2 R_y(A_2) + \frac{17}{12} g_Y^2\bigg)  \\
(4\pi)^2 \frac{d\lambda_H}{d\ln\mu} &=& 24 \lambda^2_H +\lambda_H  \left(12 y_t^2 -9 g_2^2 
R_\lambda(A_2) 
-3 g_Y^2\right) + \nonumber\\
&&+\frac{9 g_2^4}{8} R_g(A_2)+\frac{3 g_Y^4}{8}+\frac{3 g_2^2 g_Y^2}{4}R_g(A_2,0) -6 y_t^4.
\eea
We provide a numerical example in fig.\fig{GaugeRunning} (right).
Once $g_2$ approaches its fixed point, $y_t$ runs in a way slightly different way than in the SM:
it can become larger or smaller depending on the fixed-point value of $g_2$.
More importantly, the negative value of $R_g(1) \simeq -2/9$ together with the enhanced $g_2$
makes the Higgs quartic more negative at large energies, conflicting with bounds from vacuum meta-stability,
$\lambda_H \circa{>}-0.05$.  The conflict is reduced by keeping $g_2$ as small as in the SM.
However, in order to avoid this problem, one needs to extend the SM in a way that avoids vacuum instability.
The simplest option is adding one extra scalar that gives a tree-level positive correction to $\lambda_H$~\cite{1203.0237}.

\subsection{Asymptotically safe $\U(1)_Y$?}
Since $\SU(2)_L$ and $\SU(3)_c$ in the SM are anyhow asymptotically free, while  hypercharge has
a possible Landau pole around $10^{40}\GeV$, it would be especially interesting to bypass it by making
hypercharge asymptotically safe.
We add to the SM $N_F$ fermions with hypercharge $\pm Y$ and singlet under
$\SU(2)_L$ and $\SU(3)_c$ 
such that $\Delta b_Y = 4N_F Y^2 /3$, while $\Delta b_3 = \Delta b_2=0$.
Then, the resummed one-loop beta functions for the top Yukawa coupling and for
the Higgs quartic $\lambda_H$ are
\bea
(4\pi)^2 \frac{dy_t}{d\ln\mu} &=& \frac92 y_t^3 - y_t \bigg(8 g_3^2 + \frac94 g_2^2  + \frac{17}{12} g_Y^2 R_y(A_Y)\bigg)  \\
(4\pi)^2 \frac{d\lambda_H}{d\ln\mu} &=& 24 \lambda^2_H +\lambda_H  \left(12 y_t^2 -9 g_2^2 
-3 g_Y^2 R_\lambda(A_Y) \right) + \nonumber\\
&&+\frac{9 g_2^4}{8} +\frac{3 g_Y^4}{8}R_g(A_Y)+\frac{3 g_2^2 g_Y^2}{4}R_g(A_Y,0) -6 y_t^4
\eea
The fixed point for $g_Y$ corresponds to
\beq 1 - \frac{2}{5} A_Y \simeq \exp\bigg[-\frac{45}{28\pi^2}(b_Y^{\rm SM} + \Delta b_Y + 0.4)\bigg]\eeq
showing that  it is exponentially close to the pole at $A_Y \simeq 5/2$.
The functions $R_y$, $R_g$, $R_\lambda$ too have poles at $A_Y=5/2$.
As a result, when $g_Y$ approaches its fixed point, 
$y_t$ is driven to small values~\cite{1712.06859} and $\lambda_H$ is driven to
non-perturbatively large values.
Thereby it is not possible to make the Abelian factor asymptotically safe.

\medskip

Even adding $N_F$ fermions charged under all the SM group factors,
the problems related to the Abelian factor prevents us from building a full asymptotically
safe extension of the SM.
This can be build embedding the SM in non-Abelian groups, even at the weak scale, similarly
to what already done for asymptotically free extension~\cite{1412.2769,1507.06848}.

\section{Conclusions}\label{concl}
Theories where all couplings can be extrapolated up to infinite energy are interesting per se, and offer alternative solutions
to the Higgs mass hierarchy problem.
However, in the SM, the hypercharge gauge coupling grows with energy. 
Naively, adding a large number $N_F$ of  extra charged fermions goes in the wrong direction,
as  the hypercharge  coupling grows even faster than in the SM. 
Fortunately, the very large number of fermion limit helps  to tame the high energy growth of the coupling. The leading contribution of the large number of fermions can be resummed.

By computing all the resummed RGE at leading order in $1/N_F$, we found that the non-Abelian factors of the SM gauge group can be made asymptotically safe.  However, when hypercharge is made asymptotically safe, the Higgs quartic flows out of perturbative control.

\medskip
It should  be noted that 
the large $N_F$ limit is  merely a mathematical tool that allows us to determine the location of the asymptotically safe fixed points.
Lattice simulations may very well find that fixed points exist  for moderate values of $N_F$ ---
 after all, large-$N_c$ approximations are used in QCD where $N_c=3$. 
 In the meantime, it is interesting to discuss the unusual physics resulting from having many extra degrees of freedom.
 Electroweak corrections to the $W$ precision parameter get enhanced~\cite{hep-ph/0405040,1609.08157},
 tails of $d\sigma(pp\to \ell^+\ell^-)/dm_{\ell^+\ell^-}$ at large invariant mass would exhibit the pattern typical of fast running $g_{2}$ coupling~\cite{1410.6810,1602.04801}. 
 For  extra colored vector-like fermions, the modified high energy behaviour would affect the three to two jet ratio~\cite{1403.7411}. 
More interestingly, freeze-out of a keV-scale relativistic sterile neutrino from a plasma with $N_F\gg 1$ extra degrees of freedom
provides an acceptable cold Dark Matter candidate
(rather than the usual too warm DM). 
Furthermore, one can gauge the $\SU(N_F)$ symmetry that rotates the $N_F$ fermions,
that can be identified as `dark baryons' in models of composite Dark Matter.
A new feature of  $N_F\gg 1$ is that the model is phenomenologically acceptable even when stable
dark baryons are charged:
their charge grows with $N_F$, but their relic abundance gets suppressed by $2^{-N_F}$.
This discussion exemplifies the new spectrum of possibilities with atypical phenomenology
in (astro)particle physics and cosmology that these constructions  open up.

\appendix

\section{Resummed gauge corrections}
We consider a simple gauge group $G$ with gauge coupling $g$ and $N_F\gg 1$ fermions $\psi_j$ in a generic representation 
of $G$. 
We here compute the $\beta$-functions of the scalar quartics  and Yukawa couplings with the gauge field propagator 
obtained by resumming the effects of the fermions $\psi_j$ at the leading order in $1/N_F$. 

We define $A \equiv \Delta b \,  \alpha/4\pi$,
where   $\alpha \equiv g^2/4\pi$ and $\Delta b$ is the contribution of the $N_F$ fermions to the
one-loop coefficient of the gauge $\beta$-function,
$\beta_{\alpha}^{\rm one-loop} = \Delta b\, \sfrac{\alpha^2 }{2\pi}$.
The  limit $N_F \to \infty$ is taken by keeping $A$ fixed.

In order to perform the resummation of the leading terms in the expansion in $1/N_F$ we use the resummed gauge field propagator ${\mathscr{D}_{\mu\nu}}(k)$, where $k$ is the momentum. 
We choose the Landau gauge, where the tree-level propagator ${D}_{\mu\nu}(k)$ is transverse.
      \be D_{\mu\nu}(k) =-i \frac{ P_{\mu\nu}(k)}{k^2}, \qquad  
      P_{\mu\nu}(k) = \eta_{\mu\nu}- \frac{k_\mu k_\nu}{k^2}.
      \ee
The Feynman $i\varepsilon$ is left implicit.
In this  gauge the resummed propagator  ${\mathscr{D}_{\mu\nu}}(k)$ is
       \be {\mathscr D}_{\mu\nu}(k) =-i \frac{ P_{\mu\nu}(k)}{k^2} \sum_{n=0}^\infty \Pi(k^2)^n =-i  \frac{ P_{\mu\nu}(k)}{k^2} \frac1{1-\Pi(k^2)} 
       \label{resummedPi}
      \ee
   where $\Pi(k^2)$ is defined in terms of the correction to the
  vector self-energy $\Pi_{\mu\nu}(k)$   due to a loop of the $N_F$ fermions as follows
   \be \Pi_{\mu\nu}(k) = k^2P_{\mu\nu}(k) \Pi(k^2).\ee
    In dimensional regularization ($d=4-\epsilon$) one finds~\cite{1712.06859}
\be\Pi(k^2) =\left(\frac{\mu^2}{k^2}\right)^{\epsilon/2}  \Pi_0 , \qquad \mbox{with} \qquad \Pi_0 \equiv -6(-4\pi)^{\epsilon/2} A_0 \frac{\Gamma\left(\sfrac{\epsilon}{2}\right)\Gamma\left(2-\sfrac{\epsilon}{2}\right)^2}{\Gamma\left(4-\epsilon\right)}, \label{Pi}\ee 
where $A_0$ is the bare value of $A$.

\subsection{Quartic $\beta$-function}\label{Rlambda}
We here compute the $\beta$-functions of the scalar quartic couplings. 
We consider a set of real scalars $\phi_a$ in a generic representation $S$ of $G$. 
The covariant kinetic terms of the $\phi_a$ appear in Lagrangian as $\frac12 D_\mu \phi_a D^\mu \phi_a$, where the scalar covariant derivative is given by $D_\mu \phi_a = (\partial_\mu \phi_a+ i g \theta^B_{ab} A^B_\mu \phi_b)$ and the $\theta^A$ are the generators in the representation $S$.  We write their quartic interactions in the Lagrangian as $-\lambda_{abcd}\phi_a\phi_b\phi_c\phi_d/4!$. Here we show that the $\lambda_{abcd}$ obey the following RGEs at leading order in $1/N_F$:
\be (4\pi)^2 \frac{d\lambda_{abcd}}{d\ln \mu} =6 g^4 R_g(A) \theta_{abcd}   -3 g^2 R_\lambda(A)  \lambda_{abcd}  \sum_{k=a,b,c,d} C_{H}^k + {\cal O}(\lambda^2, \lambda y^2, y^4), \label{betalambdaabcd}
 \ee
 where
\be\theta_{abcd} \equiv  \frac1{16}\sum_{\rm perms}   \{\theta^A,\theta^B\}_{ab}\{\theta^A ,\theta^B\}_{cd}. \ee
The sum runs over all permutations of $abcd$, 
 the $C_{H}^{a}$ are defined by $\theta^A_{ac} \theta^A_{cb}= C_{H}^{a} \delta_{ab}$
 and the functions $R_g(A) $ and $R_\lambda(A)$ are given in eq.s~(\ref{Rgeq}) and~(\ref{Rlambdaeq}).
  For $A\approx 0$ one has $R_g= 1 -8A/3+{\cal O}(A^2)$ and $R_\lambda = 1-5A/6+{\cal O}(A^2)$ in agreement with known 2-loop results~\cite{Machacek:1984zw}.
 
 \begin{figure}[t]
\begin{center}
\includegraphics[width=0.9\textwidth]{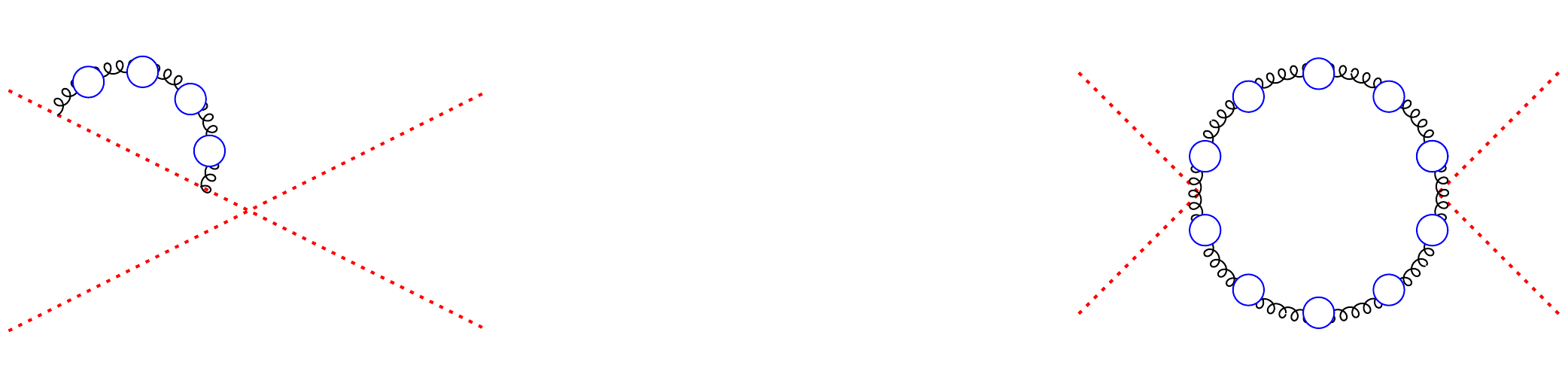}
\caption{\label{quartic1}\em Contributions at leading order in $1/N_F$ to the $\beta$-functions of the quartic couplings from the scalar field renormalizations   (left)  and the vertex contribution (right).
 }
\end{center}
\end{figure}

\subsubsection{Scalar field renormalization and $R_\lambda$}\label{Scalar field renormalization}
In Landau gauge $R_\lambda$  is given only by the scalar field renormalizations
(diagram in fig.~\ref{quartic1}, left). 
We define $Z_k$ through $\phi_{0k} = Z_k^{1/2} \phi_{k}$, where the $\phi_{0k}$ are the bare scalar fields.
The (amputated) Feynman amplitude for the scalar field renormalization of the $k^{\rm th}$ external line with the insertion of $n$ bubbles is denoted as $-i S^{(n)}_k(p^2)$, where $p$ is the momentum flowing through the diagram. 
Their resummation is $S_k\equiv \sum_n S_k^{(n)}$.
We have 
\be Z_k=1 + \left.\frac{dS_k(p^2)}{dp^2}\right|_{\rm poles}\label{polestoZ}.\ee
A calculation of $Z_k$ in the Landau gauge was presented in~\cite{1712.06859}; however, we find a different result and, therefore, we provide here the details of the calculation.
 By using eq.~(\ref{resummedPi}) and (\ref{Pi}) we find 
 \be \frac{dS_k(p^2)}{dp^2}  = - \frac{2C_{H}^k}{\Delta b} 
  \sum_{n=1}^\infty\left(-2 A_0\right)^{n} \frac{A_\phi(n,\epsilon)}{n\epsilon^{n}}. \label{Sexp1} \ee 
We do not provide the explicit expression for $A_\phi(n,\epsilon)$ for $n>0$ because it is complicated and, as we will see, the only quantity we  need is $A_\phi(0,\epsilon)$. 
Following~\cite{PalanquesMestre:1983zy} we  expand $A_\phi(n,\epsilon)$ as
\be A_\phi(n,\epsilon)= \sum_{j=0}^{\infty}A_\phi^{(j)}(\epsilon)(n\epsilon)^j.\label{ExpEN}\ee 
One can explicitly check that the coefficients $A_\phi^{(j)}(\epsilon)$ do not have poles at $\epsilon=0$.
By inserting the expansion eq.~(\ref{ExpEN}) in (\ref{Sexp1})  
 one obtains
\be  \frac{dS_k(p^2)}{dp^2}
 = -\frac{2C_{H}^k}{\Delta b} \sum_{n=1}^\infty\left(-2 A_0\right)^n \sum_{j=0}^{\infty}A_\phi^{(j)}(\epsilon)\frac{n^{j-1}}{\epsilon^{n-j}}. \label{Sexp2}\ee 
In order to obtain the $\beta$-function we express the poles in terms of the renormalized couplings. In the case of the gauge coupling this consists in using the relation  
\be A_0 = \frac{A}{Z_3} \qquad \mbox{where } \qquad Z_3 =1-\frac{2A}{\epsilon}  +{\cal O}(\frac{1}{N_F})  .\label{K0K}\ee
Then, by using the binomial series we obtain
\be A_0^n = A^n \sum_{i=0}^\infty \left(\bac -n \\
i \ea \right) \left(-\frac{2A}{\epsilon}\right)^i +{\cal O}(\frac{1}{N_F})  \ee 
which, inserted in eq.~(\ref{Sexp2}), gives (dropping sub-leading powers in $1/N_F$)
\bea \frac{dS_k(p^2)}{dp^2}   &=& 
-\frac{2C_{H}^k}{\Delta b}\sum_{n=1}^\infty\left(-2A\right)^n \sum_{j=0}^{\infty}A_\phi^{(j)}(\epsilon)\frac{n^{j-1}}{\epsilon^{n-j}}\sum_{i=0}^\infty \left(\bac -n \\
i \ea \right) \left(-\frac{2A}{\epsilon}\right)^i \nonumber \\
&=& - \frac{2C_{H}^k}{\Delta b}\sum_{n=1}^\infty\left(-2A\right)^n \sum_{j=0}^{\infty}A_\phi^{(j)}(\epsilon)\frac{n^{j-1}}{\epsilon^{n-j}}\sum_{i=0}^\infty (-1)^i \left(\bac n+i-1 \\
i \ea \right) \left(-\frac{2A}{\epsilon}\right)^i \nonumber\\
&=& - \frac{2C_{H}^k}{\Delta b} \sum_{n=1}^\infty\left(-2A\right)^n\sum_{j=0}^{\infty}\frac{A_\phi^{(j)}(\epsilon)}{\epsilon^{n-j}} \sum_{i=0}^{n-1}(n-i)^{j-1} (-1)^i \left(\bac n-1 \\
i \ea \right)  .
 \label{Sexp3}\eea
In the last step, we substituted $n\to n-i$, which requires the sum over $i$ to be truncated at $i=n-1$. 
 Keeping only the poles at $\epsilon =0$, which are the only thing we need to compute $Z_k$ (see eq.~(\ref{polestoZ})), we obtain 
\be \left.\frac{dS_k(p^2)}{dp^2}\right|_{\rm poles}     = - \frac{2C_{H}^k}{\Delta b} \sum_{n=1}^\infty\left(-2A\right)^n\sum_{j=0}^{n-1}\frac{A_\phi^{(j)}(\epsilon)}{\epsilon^{n-j}} \sum_{i=0}^{n-1}(n-i)^{j-1} (-1)^i \left(\bac n-1 \\
i \ea \right).\ee
 The sum over $i$ in the last expression is~\cite{PalanquesMestre:1983zy}
 \be \sum_{i=0}^{n-1}(n-i)^{j-1} (-1)^i \left(\bac n-1 \\
i \ea \right)  = -  \delta_{j0} \frac{(-1)^n}{n}  \qquad (\mbox{for} \, \, 0\leq j \leq n-1) \label{magic}
\ee
so only the term with $j=0$ matters and we find
\be \left.\frac{dS_k(p^2)}{dp^2} \right|_{\rm poles}     = \frac{2C_{H}^k}{\Delta b} \sum_{n=1}^\infty\left(2A\right)^n\frac{A_\phi^{(0)}(\epsilon)}{n\epsilon^{n}}.\label{Sexp4}\ee
As anticipated before, only $A_\phi^{(0)}(\epsilon) \equiv A_\phi(0,\epsilon)$ is relevant. We have
\be A_\phi(0,\epsilon)= \frac{2^{2-\epsilon } (1-\frac{\epsilon}{3}) \Gamma \left(\frac{5}{2}-\frac{\epsilon }{2}\right)}{ \sqrt{\pi } (1-\frac{\epsilon}{2}) \Gamma \left(1-\frac{\epsilon }{2}\right) \Gamma \left(3-\frac{\epsilon }{2}\right) \Gamma \left(\frac{\epsilon }{2}+1\right)}.
 \label{Aphi0}\ee
 This expression agrees with the known 2-loop expressions,
 and  does not agree with the corresponding expression in~\cite{1712.06859}
that computed the resummed Yukawa  $\beta$-function (on which we will agree).

In order to compute the correction $R_\lambda$ to the quartic  $\beta$-function we only need 
the simple pole at $\epsilon=0$. 
Using that $A_\phi(0,\epsilon)$ is regular in $\epsilon=0$ we expand
\be A_\phi^{(0)}(\epsilon) = \sum_{i=0}^\infty\tilde A_\phi^{(i)} \epsilon^i, \qquad \mbox{where} \qquad  \tilde A_\phi^{(i)} =\frac1{i!} \frac{d^iA_\phi^{(0)}}{d\epsilon^i} (\epsilon=0) \ee
which, once inserted in (\ref{Sexp4}), gives the simple pole
\be \left.\frac{dS_k(p^2)}{dp^2}\right|_{\rm simple \,\, pole}   =  \frac{2C_{H}^k}{\Delta b} \frac1{\epsilon} \sum_{n=1}^\infty\left(2A\right)^n\frac{\tilde A_\phi^{(n-1)}}{n}.\label{Sexp5}\ee

This result allows us to compute the term proportional to $g^2\lambda_{abcd}$ in the $\beta$-function of $\lambda_{abcd}$. Using the general formula provided in~\cite{Weinberg:1996kr} one sees that the $\beta$-function in this case is obtained by taking a logarithmic derivative with respect to $A$ of (\ref{Sexp5}). Then, the $n$ in the denominator of (\ref{Sexp5}) disappears  and one finds a closed form for the $\beta$-function. The result is given in eq.s~(\ref{betalambdaabcd}) and~(\ref{Rlambdaeq}).  

\subsubsection{Correction to the quartic vertex and $R_g$}
Next, we compute $R_g$ in eq.~(\ref{betalambdaabcd}) by considering 
 the four-point loop vertex  at the leading order in $1/N_F$ at vanishing external momenta. 
 In the Landau gauge only the diagram in the right panel of fig.~\ref{quartic1} contributes.
The relative Feynman amplitude $iV_{abcd}$ is given by
\be \frac{V_{abcd}}{\mu^\epsilon} = -4i \mu^\epsilon \theta_{abcd} g_0^4 \sum_{n,m = 0 }^\infty (\Pi_0 \mu^\epsilon)^{n+m} \int \frac{d^d q}{(2\pi)^d} \frac{d-1}{(q^2 -m_\gamma^2)^2 (q^2)^{(n+m)\epsilon/2}}, \label{V4p} \ee 
where $m_\gamma$ is a tiny vector mass added to regulate IR divergences.  The $\beta$-function of $\lambda_{abcd}$ is insensitive to this divergence and so we set $m_\gamma \to 0$ at the end.
Eq.~(\ref{V4p}) has been obtained by substituting in the relevant one-loop diagram the tree-level gauge field propagator $D_{\mu\nu}$ with the resummed propagator $\mathscr{D}_{\mu\nu}$ of eq.~(\ref{resummedPi}). The double sum over $n$ and $m$ is due to the presence of two gauge field propagators in the diagram.
Computing  the loop integrals in eq.~(\ref{V4p}) gives
 \be \frac{V_{abcd}}{\mu^\epsilon} = \frac{16\theta_{abcd}}{9\Delta b^2} \sum_{n=2}^{\infty} \left(-2A_0\right)^{n} \frac{A_4(n,\epsilon)}{n \epsilon^{n-1}}   \ee
 for some coefficients $A_4(n,\epsilon)$. 
 We do not display  $A_4(n,\epsilon)$ because it is complicated for generic $n$,
 and because  we only need  $A_4(0,\epsilon)$ and $A_4(1,\epsilon)$.
 The double sum in eq.~(\ref{V4p}) has been reduced to a single sum through the identity (valid for any sequence $c_n$)
 \be \sum_{n,m = 0}^\infty c_{n+m} =  \sum_{n=0}^\infty (n+1) c_n. \ee 
We then follow an approach similar to the one used for the scalar field renormalization. We expand
\be A_4(n,\epsilon)= \sum_{j=0}^{\infty}A_4^{(j)}(\epsilon)(n\epsilon)^j,\label{ExpEN4}\ee 
which, once inserted in the expression above for $V_4$, gives
 \be\frac{V_{abcd}}{\mu^\epsilon} = \frac{16\theta_{abcd}}{9\Delta b^2} \sum_{n=2}^{\infty} \left(-2A_0\right)^{n} \sum_{j=0}^\infty  A_4^{(j)}(\epsilon) \frac{n^{j-1}}{\epsilon^{n-1-j}}. 
 \ee
 By using again the relation between the bare $A_0$ and the renormalized $A$, eq.~(\ref{K0K}), and performing steps similar to those done for the scalar field renormalization one finds
 \be \frac{V_{abcd}}{\mu^\epsilon} = \frac{16\theta_{abcd}}{9\Delta b^2}\sum_{n=2}^{\infty} \left( -2A_0\right)^{n} \sum_{j=0}^\infty  A_4^{(j)}(\epsilon) \frac{n^{j-1}}{\epsilon^{n-1-j}}  \sum_{i=0}^\infty (-1)^i \left(\bac n+i-1 \\
i \ea \right) \left(-\frac{2A}{\epsilon}\right)^i. \ee
We now replace $n\to n-i$, which requires here to stop the sum over $i$ at $i=n-2$, so
 \be\frac{V_{abcd}}{\mu^\epsilon} = \frac{16\theta_{abcd}}{9\Delta b^2}\sum_{n=2}^{\infty} \left(-2A\right)^{n} \sum_{j=0}^\infty   \frac{A_4^{(j)}(\epsilon)}{\epsilon^{n-1-j}}  \sum_{i=0}^{n-2}(n-i)^{j-1} (-1)^i \left(\bac n-1 \\
i \ea \right). \ee
Next, we rewrite the sum over $i$ as 
\be  \sum_{i=0}^{n-2}(n-i)^{j-1} (-1)^i \left(\bac n-1 \\
i \ea \right) = (-1)^n + \sum_{i=0}^{n-1}(n-i)^{j-1} (-1)^i \left(\bac n-1 \\
i \ea \right),\ee 
which leads to
\be  \frac{V_{abcd}}{\mu^\epsilon} = \frac{16\theta_{abcd}}{9\Delta b^2}\sum_{n=2}^{\infty}\left(2A\right)^{n}\frac1{\epsilon^{n-1}} \left[ \sum_{j=0}^\infty \epsilon^j A_4^{(j)}(\epsilon) - \frac{A_4^{(0)}(\epsilon)}{n} \right],  \ee 
where we  used that $\epsilon A_4^{(j)}(\epsilon)$ are regular at $\epsilon = 0$ and, therefore, we have used eq.~(\ref{magic}) to compute the second term proportional to $A_4^{(0)}(\epsilon)$. An explicit calculation shows $A_4^{(0)}(\epsilon) =  0 $ and we are therefore left with 
 \be  \frac{V_{abcd}}{\mu^\epsilon} = \frac{16\theta_{abcd}}{9\Delta b^2} \sum_{n=2}^{\infty}\left(2A\right)^{n}\frac{A_4(1,\epsilon)}{\epsilon^{n-1}},  \ee 
having used eq.~(\ref{ExpEN4}). The explicit expression of $A_4(1,\epsilon)$ is
 \be A_4(1,\epsilon)= -\frac{3 \pi ^2 (\epsilon -3) \Gamma (4-\epsilon )}{\Gamma \left(2-\sfrac{\epsilon }{2}\right)^3 \Gamma \left(\sfrac{\epsilon }{2}+1\right)}. \label{A41}\ee
  In order to compute the $\beta$-function we only need the simple pole, so we expand 
 \be A_4(1,\epsilon) = \sum_{i = 0 }^\infty \tilde A_4^{(i)} \epsilon^i \label{A4exp}\ee 
 and insert this expansion in the last expression of $V_{abcd}$, to obtain
  \be  \left.\frac{V_{abcd}}{\mu^\epsilon} \right|_{\rm simple \, pole}=  \frac{16\theta_{abcd}}{9\Delta b^2\epsilon} \sum_{n=2}^{\infty}\left(2A\right)^{n} \tilde A_4^{(n-2)}  
  =\frac{64\theta_{abcd}A^2}{9\Delta b^2\epsilon}  A_4(1, 2A). \label{spV41}\ee
 Using again the general formula provided in~\cite{Weinberg:1996kr} (which allows us to extract the $\beta$-function from the simple pole)  leads to the result in eq.~(\ref{betalambdaabcd}) and~(\ref{Rgeq}).

\medskip

  It is worth noting that one can perform these calculations  by using directly the resummed gauge field propagator in the last equality in eq.~(\ref{resummedPi}) before doing the loop integral. Let us illustrate this method in the calculation of $V_{abcd}$. In this case
\be \frac{V_{abcd}}{\mu^\epsilon} = -4i \mu^\epsilon \theta_{abcd} g_0^4 \int \frac{d^d q}{(2\pi)^d} \frac{d-1}{  (q^2-q^2\Pi(q^2))^2}. \label{V4} \ee 
Using the expression of $\Pi(k^2)$ given in eq.~(\ref{Pi}), the loop integral can be done analytically after performing the Wick rotation.
After expressing the bare $A_0$ in term of the renormalized $A$ by means of eq.~(\ref{K0K}), one obtains 
\be  \frac{V_{abcd}}{\mu^\epsilon} =  \frac{32\theta_{abcd}}{9\Delta b^2}\frac{A A_4(1,\epsilon)}{1-{2A}/{\epsilon}}, \ee
where $A_4(1,\epsilon)$ is given in (\ref{A41}). To compute the $\beta$-function we only need the simple pole in this expression. Therefore, we use the expansion in (\ref{A4exp}) and 
\be \frac1{1-\sfrac{2A}{\epsilon}} = \sum_{j=0}^\infty \left(\frac{2A}{\epsilon}\right)^j\ee 
to obtain
\be  \left.\frac{V_{abcd}}{\mu^\epsilon}\right|_{\rm simple \, pole} =   \frac{32\theta_{abcd}A}{9\Delta b^2\epsilon} \sum_{j=1}^\infty \left(2A\right)^j \tilde A_4^{(j-1)} = 
\frac{64\theta_{abcd}A^2}{9\Delta b^2\epsilon}  A_4(1,2A), \ee 
which coincides with eq.~(\ref{spV41}), obtained instead by first performing the loop integral and then   resumming. This provides another  check for the $\beta$-function of $\lambda_{abcd}$.

The derivation of $R_g(A,0)$ is not explicitly presented because it is very similar to the derivation of $R_\lambda(A)$.
   
\subsection{Yukawa $\beta$-function}\label{Ry}

We now present the derivation of the $\beta$-function of the Yukawa coupling discussed in section~\ref{YukSec}. 
A calculation of $\beta_y$ was presented in~\cite{1712.06859} and we do agree with their final expression. However, their derivation mixes the Feynman and Landau gauge in an {\it apparently} inconsistent way, so we preferred to provide another independent calculation performed in the Landau gauge. 

To compute $\beta_y$ three ingredients are required: the scalar field renormalization (already computed in section~\ref{Scalar field renormalization}), the fermion renormalization and the loop correction to the Yukawa vertex. We provide the derivation of the remaining pieces in the following part of the appendix. Once the simple poles of these three ingredients are obtained one can determine $\beta_y$ with  the general formula provided in~\cite{Weinberg:1996kr} and the result is the one given in section~\ref{YukSec}.

\subsubsection{Fermion field renormalization}
One ingredient to obtain $\beta_y$ is the fermion field renormalization. The (amputated) Feynman amplitude of the self-energy of the fermions $\psi_1$ and $\psi_2$ is 
\be -i \Sigma_{1,2} (p)=  (ig_0)^2 \mu^\epsilon C_{\psi_{1,2}}
 \int\frac{d^dk}{(2\pi)^d} \gamma_\mu
\frac{i}{\slashed{p}-\slashed{k}} \gamma_\nu \frac{-i P_{\mu\nu} (k)}{k^2(1-\Pi(k^2))},
 \ee 
where $C_{\psi_{1,2}}$  is the quadratic Casimir of $\psi_{1,2}$ under $G$.\footnote{We displayed explicitly this form of $\Sigma_{1,2}$ to make the comparison with the corresponding expression in~\cite{1712.06859} easier. While we use the Landau gauge consistently it seems that~\cite{1712.06859} simply replaced $P_{\mu\nu}$ with $\eta_{\mu\nu}$, which appears to be an inconsistent mixing of the Landau and Feynman gauge (as the resummation of the gauge field propagator, eq.~(\ref{resummedPi}), is performed in the Landau gauge both here and in~\cite{1712.06859}).}
After expanding $1/(1-\Pi)$ in a power series of $\Pi$ and dealing with $\gamma$-matrices we extract $ \sfrac{d\Sigma_{1,2}}{d\slashed{p}} $
at zero external momentum:
\be \left.\frac{d\Sigma_{1,2}}{d\slashed{p}} \right|_{p=0}= -i \mu^\epsilon g_0^2 C_{\psi_{1,2}}\frac{4+d^2-5d}{d} \int \frac{d^dk}{(2\pi)^d} \frac{1}{k^4} \sum_{n=0}^\infty \Pi(k^2)^n\ee
as this is the quantity needed to obtain the fermion renormalizations $Z_{1,2}$: 
\be Z_{1,2}-1 =\left.\frac{d\Sigma_{1,2}}{d\slashed{p}}\right|_{p=0,\rm poles}.  \ee
Once again, we rewrite the relevant expression as 
\be \left.\frac{d\Sigma_{1,2}}{d\slashed{p}} \right|_{p=0}=  \frac{C_{\psi_{1,2}}}{\Delta b} \sum_{n=1}^\infty (-2A_0)^n \frac{A_\psi(n,\epsilon)}{n\epsilon^{n}}  \ee 
for some  $A_\psi(n,\epsilon)$. Using a method similar to section~\ref{Scalar field renormalization} we find the following simple pole:  
\be \left.\frac{d\Sigma_{1,2}}{d\slashed{p}}\right|_{p=0,\rm simple \,\, pole}   =  -\frac{C_{\psi_{1,2}}}{\Delta b} \frac1{\epsilon} \sum_{n=1}^\infty\left(2A\right)^n\frac{\tilde A_\psi^{(n-1)}}{n},\label{Sexp6}\ee
where the coefficients $\tilde A^{(n-1)}_\psi$ are defined by
\be A_\psi(0,\epsilon) = \sum_{i=0}^\infty\tilde A_\psi^{(i)} \epsilon^i, \qquad \mbox{where} \qquad  \tilde A_\psi^{(i)} =\frac1{i!} \frac{d^iA_\psi^{(0)}}{d\epsilon^i} (\epsilon=0) \ee
and 
\be A_\psi(0,\epsilon) =  \frac{(3-\epsilon) \epsilon  \Gamma (4-\epsilon )}{6 (\epsilon -4) \Gamma \left(2-\sfrac{\epsilon }{2}\right)^3 \Gamma \left(1+\sfrac{\epsilon }{2}\right)}.\ee

\subsubsection{Correction to the Yukawa vertex}
The remaining ingredient to calculate $\beta_y$ is the loop correction to the Yukawa vertex, whose (amputated) Feynman amplitude is denoted here with  $-i \Lambda_y$. We can set the external momenta to zero as the Yukawa coupling is a non-derivative interaction. We obtain 
\be \frac{\Lambda_y}{\mu^\epsilon} = -\frac{6 y}{\Delta b} (C_{\psi_1}+C_{\psi_2}-C_H) \sum_{n=1}^\infty (-2A_0)^n \frac{A_y(n,\epsilon)}{n\epsilon^n} \ee 
for some $A_y(n,\epsilon)$ and, by using a technique similar section~\ref{Scalar field renormalization}, we find the  simple pole:  
\be \left.\frac{\Lambda_y}{\mu^\epsilon}\right|_{\rm simple \,\, pole} = \frac{6 y}{\Delta b} (C_{\psi_1}+C_{\psi_2}-C_H) \frac{1}{\epsilon}\sum_{n=1}^\infty (2A)^n \frac{\tilde A^{(n-1)}_y}{n} \ee
where the coefficients $\tilde A^{(n-1)}_y$ are defined by
\be A_y(0,\epsilon) = \sum_{i=0}^\infty\tilde A_y^{(i)} \epsilon^i, \qquad \mbox{where} \qquad  \tilde A_y^{(i)} =\frac1{i!} \frac{d^iA_y^{(0)}}{d\epsilon^i} (\epsilon=0) \ee
and 
\be A_y(0,\epsilon) = \frac{(3-\epsilon) \Gamma (4-\epsilon )}{24 \Gamma \left(2-\sfrac{\epsilon }{2}\right)^3 \Gamma \left(1+\sfrac{\epsilon }{2}\right)}.  \ee

\subsection*{Acknowledgements}
This work was supported by the grant 669668 -- NEO-NAT -- ERC-AdG-2014 and the Danish National Research Foundation Grant, DNRF-90. ADP acknowledges financial support from CONACyT and the NEO-NAT grant.

\footnotesize

\bibliographystyle{abbrv}
\bibliography{mybib}

\begin{thebibliography}{nnn}
\bibitem{1707.01303}
\heparticle[1707.01303]{CMS Collaboration}{Search for heavy resonances that decay into a vector boson and a Higgs boson in hadronic final states at $\sqrt{s}=13 \TeV$}.
\end{thebibliography}

\end{document}